\documentclass[a4paper,fleqn,usenatbib]{mnras}

\usepackage[T1]{fontenc}
\usepackage{ae,aecompl}

\newcommand{\dego}{$^{o}$}

\newcommand{\Msun}{\hbox{$\hbox{M}_\odot\;$}}

\newcommand{\kms}{\hbox{${\rm km}\:{\rm s}^{-1}\;$}}
\newcommand{\Msuno}{\hbox{$\hbox{M}_\odot$}}
\newcommand{\Rsuno}{\hbox{$\hbox{R}_\odot$}}
\newcommand{\kmso}{\hbox{${\rm km}\:{\rm s}^{-1}$}}

\usepackage{graphicx}
\usepackage{natbib}
\usepackage{amsmath}	
\usepackage{amssymb}	
\def\aj{AJ}%
\def\actaa{Acta Astron.}%
\def\araa{ARA\&A}%
\def\apj{ApJ}%
\def\apjl{ApJ}%
\def\apjs{ApJS}%
\def\ao{Appl.~Opt.}%
\def\aap{A\&A}%
\def\mnras{MNRAS}%
\def\na{New A}%
\def\prd{Phys.~Rev.~D}%
\def\nat{Nature}%
\title[Extremely fast orbital decay of the black hole X-ray binary Nova Muscae 1991]
{Extremely fast orbital decay of the black hole X-ray binary Nova Muscae 
1991\thanks{Based on observations taken with the X-Shooter 
spectrograph installed at the Very Large Telescope (Program ID: 091.D-0921(A)), 
at Paranal Observatory of the European Southern Observatory, Chile}}
\author[J.~I.~Gonz\'alez~Hern\'andez et al.]
{J.~I.~Gonz\'alez~Hern\'andez$^{1,2}$\thanks{E-mail:
jonay@iac.es}, L. Su\'arez-Andr\'es$^{1,2}$, R. Rebolo$^{1,2,3}$ and J. Casares$^{1,2}$\\
$^{1}$Instituto de Astrof{\'\i }sica de Canarias (IAC), 
E-38205 La Laguna, Tenerife, Spain\\
$^{2}$Depto. Astrof{\'\i }sica, Universidad de La 
Laguna (ULL), E-38206 La Laguna, Tenerife, Spain\\
$^{3}$Consejo Superior de Investigaciones 
Cient{\'\i}ficas, Spain
}

\date{Accepted 2016 September 08. Received 2016 September 08; in original form 2016 May 31}

\pubyear{2016}

\begin{document}
\label{firstpage}
\pagerange{\pageref{firstpage}--\pageref{lastpage}}
\maketitle

\begin{abstract}
We present new medium-resolution spectroscopic observations of the black 
hole X-ray binary \mbox{Nova Muscae 1991} taken with X-Shooter spectrograph 
installed at the 8.2m-VLT telescope. These observations allow us to measure
the time of inferior conjunction of the secondary star with the black hole in
this system that, together with previous measurements,
yield an orbital period decay of $\dot P=-20.7\pm12.7$~ms~yr$^{-1}$ 
($-24.5\pm15.1$~$\mu $s per orbital cycle).
This is significantly faster than those previously measured in the other black 
hole X-ray binaries A0620-00 and XTE J1118+480. 
No standard black hole X-ray binary evolutionary model is able to explain 
this extremely fast orbital decay. At this rate, the secondary star would reach the 
event horizon (as given by the Schwarzschild radius of about 32 km) in 
roughly 2.7~Myr.
This result has dramatic implications on the evolution and lifetime of black hole 
X-ray binaries.
\end{abstract}




\section{Introduction}

According to the standard theory, the evolution of black hole X-ray binaries 
(BHXBs) is dictated by angular momentum losses (AMLs), driven
by magnetic braking~\citep[MB;][]{ver81}, gravitational 
radiation~\citep[GR;][]{lan62,tay82} and mass loss~\citep[ML;][]{rap82}. 
MB is assumed to be the main mechanism responsible for 
AMLs in BHXBs with short orbital periods of several hours~\citep{rap82}, 
but an adequate expression for MB 
still remains a matter of debate~\citep{pod02a,iva06b,yun08b}. 
The measurement of period variations can provide valuable information on 
the strength of these processes.
The BHXB \mbox{Nova Muscae 1991} (GS Mus/GRS 1124--683) is a very 
interesting system because it has an orbital period slightly longer, 
10.38~hr~\citep{oro96,cas97}, than the two other BHXBs \mbox{XTE J1118+480} 
and \mbox{A0620--00}, for which an orbital decay measurement has been obtained~\citep{gon14}. 

The black hole mass of \mbox{Nova Muscae 1991} has been recently refined 
to $M_{\rm BH}=11.0^{+2.1}_{-1.4}$~\Msun~\citep{wu16} after 
previous determinations of $M _{\rm BH}=7.2\pm0.7$~\Msun~\citep{gel04} and 
$M _{\rm BH}=5.8^{+4.7}_{-2.0}$~\Msun~\citep{sha97}. 
Therefore, the BH mass is significantly larger (see Table~\ref{ttpar}) 
than in \mbox{XTE J1118+480}, $M_{\rm BH}\sim$~7.5~\Msun~\citep{kha13}, 
and \mbox{A0620--00}, $M_{\rm BH}\sim$~6.6~\Msun~\citep{can10}. 

\begin{table*}
\centering
\begin{minipage}{140mm}
\caption{Kinematical and dynamical binary parameters of 
\mbox{XTE J1118+480} and \mbox{A0620--00}}
\begin{tabular}{lcccccc}
\hline
{Parameter} & {Nova Muscae 1991}& {Ref.}$^\star$ & {XTEJ1118$+$480} & {Ref.}$^\star$ & {A0620$-$00} & {Ref.}$^\star$ \\
\hline
$v \sin i$~[\kmso] & $85.0\pm2.6$ & [1] & $96^{+3}_{-11}$ & [5] & $82\pm2$ & [9] \\
$i$~[\dego] & $43.2\pm2.7$ & [2] & $73.5\pm5.5$ & [6] & $51.0\pm0.9$ & [10] \\
$k_2$~[\kmso] & $406.8\pm2.7$ & [1] & $708.8\pm1.4$ & [7] & $435.4\pm0.5$ & [9] \\
$q=M_2/M_{\rm BH}$ & $0.079\pm0.007$ & [1] & $0.024\pm0.009$ & [8] & $0.060\pm0.004$ & [9] \\
$f(M)$~[\Msuno] & $3.02\pm0.06$ & [1]& $6.27\pm0.04$ & [7] & $2.762\pm0.009$ & [8] \\
$M_{\rm BH}$~[\Msun] & $11.0^{+2.1}_{-1.4}$ & [2] & $7.46^{+0.34}_{-0.69}$ & [8] & $6.61^{+0.23}_{-0.17}$ & [8] \\
$M_2$~[\Msuno] &  $0.89\pm0.18$ & [2] & $0.18\pm0.06$ & [8] & $0.40\pm0.01$ & [8] \\
$a_c$~[\Rsuno]  & $5.49\pm0.32$ & [2] & $2.54\pm0.06$ & [8] & $3.79\pm0.04$ & [8,11] \\
$R_2$~[\Rsuno] & $1.06\pm0.07$ & [2] & $0.34\pm0.05$ & [8] & $0.67\pm0.02$ & [8,11] \\
$P_{\rm orb,1}$~[d] & $0.432606(3)$ & [3] & $0.1699337(2)$ & [8] & $0.323014(4)$ & [12] \\
$P_{\rm orb,0}$~[d] & $0.432605(1)$ & [4] & $0.16993404(5)$ & [8] & $0.32301415(7)$ & [8] \\
$T_0$~[d] & $2448715.5869(27)$ & [4] & $2451868.8921(2)$ & [8] & $2446082.6671(5)$ & [8] \\
$\dot P_{\rm orb}$~[$s\;s^{-1}$] & $-6.56\pm4.03 \times 10^{-10}$ & [4] & $-6.01\pm1.81 \times 10^{-11}$ & [8] & $-1.90\pm0.26 \times 10^{-11}$ & [8] \\
$\dot P_{\rm orb}$~[ms~yr$^{-1}$] & $-20.7\pm12.7$ & [4] & $-1.90\pm0.57$ & [8] & $-0.60\pm0.08$ & [8] \\
$\dot P_{\rm orb}$~[$\mu$s~cycle$^{-1}$] & $-24.5\pm15.1$ & [4] & $-0.88\pm0.27$ & [8] & $-0.53\pm0.07$ & [8] \\
$\dot P_{\rm orb, MC,o}$~[ms~yr$^{-1}$] & $-21.1\pm12.7$ & [4] & $-1.98\pm0.56$ & [8] & $-0.63\pm0.08$ & [8] \\
$\dot P_{\rm orb, MC,c}$~[ms~yr$^{-1}$] & $-21.0\pm14.2$ & [4] & $-1.98\pm0.59$ & [8] & $-0.62\pm0.12$  & [8] \\
\hline
\end{tabular}
{\\
$^\star$~References: [1]~\citet{wu15}; [2]~\citet{wu16}; [3]~\citet{oro96}; [4]~This work; 
[5]~\citet{cal09}; [6]~\citet{kha13}; [7]~\citet{gon08b}; [8]~\citet{gon14}; [9]~\citet{nei08};
[10]~\citet{can10}; [11]~\citet{gon11}; [12]~\citet{mcc86}
}
\end{minipage}
\label{ttpar}
\end{table*}

The standard theory of the evolution of LMXBs~\citep[e.g.][]{ver93,pod02a,tay82}, predicts 
an orbital period first derivative from AMLs due to MB and ML in short-period (SP-) BHXBs 
as small as $\dot P_{\rm MB,ML} \sim -0.02$~ms~yr$^{-1}$ whereas GR accounts 
only for $\dot P_{\rm GR} \le -0.01$~ms~yr$^{-1}$, according to the dynamical 
parameters of SP-BHXBs. Recently, \citet{gon12a,gon14} reported the first detection of
 orbital period variations in two SP-BHXB and found them to be significantly larger
than expected from conventional AML theory. The period derivative measured are
$\dot P = -1.9 \pm 0.6$~ms~yr$^{-1}$ for \mbox{XTE J1118+480} and 
$\dot P = -0.6 \pm 0.1$~ms~yr$^{-1}$ for \mbox{A0620-00}. 
Extremely high magnetic fields in the secondary star at about 10--30~kG might 
explain the fast spiral-in of the companion star. Alternatively, 
unknown processes or non-standard theories of gravity have been 
suggested~\citep[e.g.][]{yag12}.
On the other hand, the detection of middle-infrared excesses from 
$Spitzer$~\citep{mun06} and WISE~\citep{wan14} have been interpreted as 
evidence for the presence of candidate circumbinary discs, or jets~\citep{gal07}, 
in these two BHXB systems. 
However, \citet{che15} have presented AML models including circumbinary discs but 
found that both the required mass transfer rate and circumbinary disc mass are far
greater than the values inferred from observations. 
This makes unlikely that circumbinary discs are the main cause of the rapid orbital 
decay observed in these two BHXBs.

\citet{gon14} suggested that the observed fast orbital decays in \mbox{XTE J1118+480} 
and  \mbox{A0620-00} could show an evolutionary sequence where the orbital period 
decay begins to speed up as the orbital period decreases.
In this work we present the detection of an extremely fast orbital period decay in the 
SP-BHXB \mbox{Nova Muscae 1991}, which in fact has a longer orbital period.

\section{Observations}

\begin{figure}
\centering
\includegraphics[height=9.2cm,angle=90]{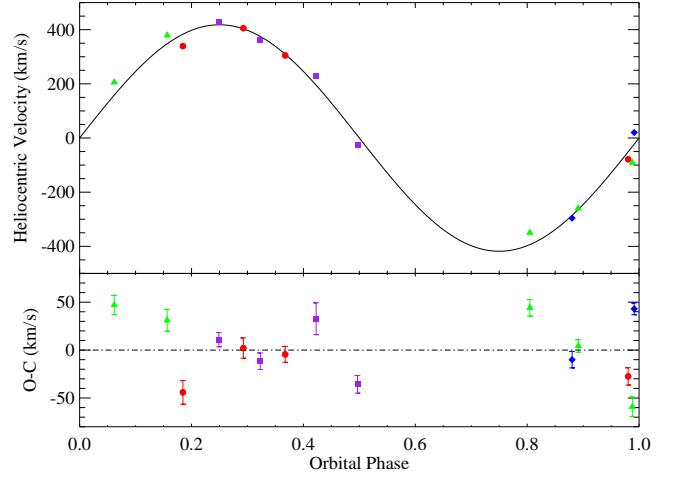}
\caption{
\scriptsize{{\it Top panel}: radial velocities of the 
secondary star in \mbox{Nova Muscae 1991} 
obtained from the VLT/X-Shooter  spectroscopic data taken 
on the four nights of  April 13 (red filled circles), April 15 (blue filled 
diamonds), and April 28 (green filled triangles), and May 09 (violet filled
squares) in 2013, folded on the best-fitting orbital solution. 
{\it Bottom panel}: residuals of the fit, with an rms of $\sim 30 {\rm km}\ {\rm s}^{-1}$.}
}
\label{frv}   
\end{figure}

\begin{figure}
\centering
\includegraphics[height=8.7cm,angle=90]{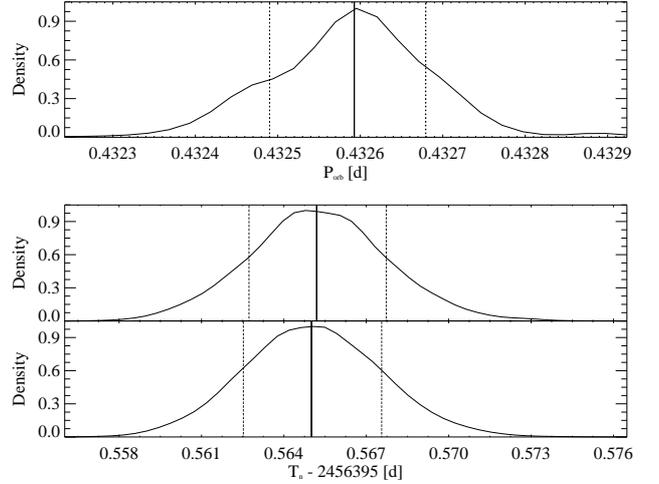}
\caption{
\scriptsize{
{\it Top panel}: posterior distribution of the orbital period, 
$P_{\rm orb}$, extracted from the MCMC Bayesian analysis of the 
new X-Shooter RV points of the secondary star in \mbox{Nova Muscae 1991}.
{\it Middle and bottom panel}: posterior distributions of the time at the inferior 
conjunction of the secondary star, $T_n$, by fixing $P_{\rm orb}$ to the 
MCMC result (middle) and by fixing $P_{\rm orb}$ to the value (bottom) 
derived by \citet{oro96}.}
}
\label{fmcmc}   
\end{figure}

We have conducted new spectroscopic observations of 
\mbox{Nova Muscae 1991} using the 8.2m VLT telescope 
equipped with the X-Shooter spectrograph~\citep{ver11} at Paranal 
Observatory [European Southern Observatory (ESO), Chile]. 
17 medium-resolution spectra 
($\lambda/\delta\lambda\sim8,800$ in the VIS arm)   
were obtained during four nights on 2013 April 13, 15, and 28 UT, and 
2013 May 09 UT. We also observed the K-dwarf template star BD-05 3763.
Radial velocity (RV) measurements (see Fig.~\ref{frv}) were extracted from 
every spectrum as in~\citet{gon12a}, by cross-correlating the observed
spectra of \mbox{Nova Muscae 1991} ($m_V\sim20.5$ at quiescence) 
with the K-dwarf template spectrum 
properly broadened with $v \sin i=85$~\kms~\citep{wu15}. 
Two RV points were rejected due to the low quality of the spectra. 
We try to correct for possible instrumental drifts during the observations
using the telluric spectra within the X-Shooter optical spectral range with 
corrections in the range 0-14 km/s.
We depict the RV points together with a Keplerian RV curve with an
orbital period fixed to the value $P_{\rm orb}=0.4326058$~d~\citep{oro96}.
This single fit to the RV points provides a value of the time at inferior 
conjunction of the secondary star, $T_n$ (see Table~\ref{ttnnmus}).
In order to evaluate the uncertainties of this $T_n$ value, we have developed
a MCMC Bayesian analysis~\citep{rub03}. We run 20 Markov chains of
75,000 iterations each resulting in a final chain of 90,000 sets of global-fit 
parameters. We first run the MCMC
code using a Keplerian function with four free parameters (including the orbital 
period). The result of the period free (P-free) simulation is shown in the upper 
panel of Fig.~\ref{fmcmc}. This orbital period determination, 
$P_{\rm orb}=0.43260(9)$ d, is consistent but less accurate 
than the orbital period derived by~\citet{oro96}. We then run two additional
period-fixed (P-fix) MCMC simulations whose result is depicted in middle
and bottom panels of Fig.~\ref{fmcmc}, one with the period fixed to the
 P-free MCMC result (middle) and the other one with the period fixed to
the orbital period derived by~\citet{oro96}. Clearly, the $T_n$ determination
do not depend on the orbital period choice to be fixed within the error bars. 
The latter becomes our final adopted value and uncertainty of the time at 
the inferior conjunction of the secondary star.

In Table~\ref{ttpar} we list the updated kinematical and dynamical
parameters of this and the two previously mentioned BHXBs.

\begin{table}
\centering
\begin{minipage}{80mm}
\caption{Time at inferior conjunction of the secondary star 
in \mbox{Nova Muscae 1991}.}
\begin{tabular}{@{}lccc@{}}
\hline
$N$ & $T_n-2440000$\footnote{Times in HJD of the $n$th 
inferior conjunction, $T_n$, of \mbox{Nova Muscae 1991}, and 
uncertainties, $\delta T_n$.} 
& $\delta T_n$ & Refs.\footnote{[1]~\citet{rem92}; [2]~\citet{oro96}; 
[3]~\citet{cas97}; [4]~\citet{wu15};  
[5]~This work: value extracted from single keplerian fit with the 
orbital period fixed (P-fix) to the value given in \citet{oro96}; 
[6]~This work: value extracted from the P-fix MCMC bayesian 
analysis by fixing the orbital period to the P-free MCMC result;
[7]~This work: value extracted from the P-fix MCMC bayesian
analysis by fixing the orbital period to the value given in \citet{oro96};
The $T_n$ values in~\citet{rem92} and~\citet{wu15} have 
been corrected from times at maximum velocity to times at orbital phase 
0 using the orbital period in these papers.} \\
\hline
0         &  8715.5875   & 0.0029  & [1] \\
0         &  8715.5888   & 0.0012  & [2] \\
2536   &  9812.6692   & 0.0010  & [3] \\
14404 & 14946.7946  & 0.0004  & [4] \\
17753 & 16395.5649  & 0.0010  & [5] \\
17753 & 16395.5652  & 0.0025  & [6] \\
17753 & 16395.5650  & 0.0025  & [7] \\
\hline
\end{tabular}
\end{minipage}
\label{ttnnmus}
\end{table}

\section{Orbital period decay\label{secdec}}

For the determination of the orbital decay we choose the 
most conservative $T_n$ value extracted from P-fix MCMC 
simulation adopting the orbital period by \citet{oro96} 
(see Table~\ref{ttnnmus}). 

Assuming a constant rate of change of the orbital period, the time 
of the $n$th orbital cycle can be expressed 
as $T_n=T_0+P_0 n+\frac{1}{2} P_0 \dot P n^2$, where $P_0$ is the 
orbital period at time $T_0$ of the reference cycle ($n=0$), 
$\dot P$ is the orbital period time derivative, and $n$, 
the orbital cycle number. 
 
We use the IDL routine {\scshape curvefit} 
to perform a parabolic fit ($\chi_\nu^2\sim7.2$) and obtain a 
period derivative of $\dot P = -(6.56 \pm1.49) \times 10^{-10}$~s/s. 
However, since the $\chi^2$ of the fit is quite large, we decided 
to multiply the error bars by a factor of 2.7, providing $\chi_\nu^2\sim1$. 
A linear function ($\dot P =0$; $\chi_\nu^2 \sim 1.5$) gives a  
worse fit, whereas a third-order polynomial 
(including $\ddot P$; $\chi_\nu^2 \sim 0.02$) gives a significantly 
lower $\chi_\nu^2$ value, although due to the still 
low number of points we do not consider this fit reliable.

We perform an F-test to evaluate how well the parabolic
fit reproduces this set of data with respect to the linear fit.
and obtain significance values of 0.07 for the first-order versus 
second-order polynomial, indicating that the second-order polynomial 
provides a better representation of the current set of data. 
 
We may note here that adding our $T_n$ value to literature values 
increases considerably the significance of the second-order polynomial 
fit with respect to the linear fit, due to the longer time baseline of the data.
Without our measurement, the significance of the linear fit grows up 
to 0.45 with respect to the second-order polynomial fit that still provides 
better representation of the data.

\begin{figure}
\centering
\includegraphics[height=9.8cm,angle=90]{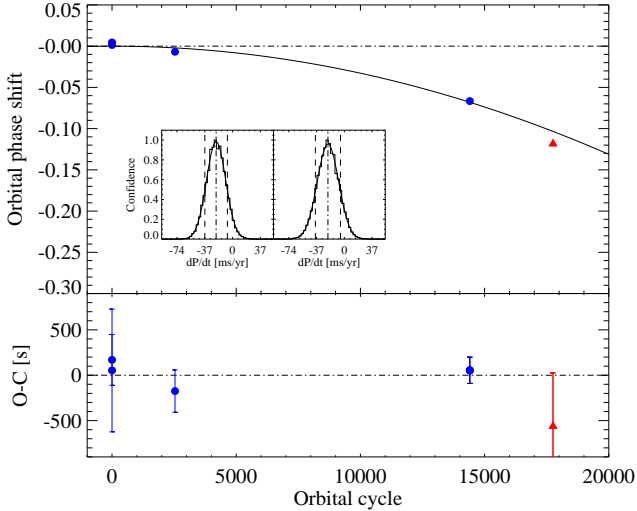}
\caption{
\scriptsize{{\it Top panel}: orbital phase shift at the time of the 
inferior conjunction (orbital phase 0), $T_n$, of the secondary star in the 
BHXB \mbox{Nova Muscae 1991} versus the orbital 
cycle number, $n$, folded on the best-fitting parabolic fit. 
The error bars give the uncertainties $\delta T_n$ (see
text and Table~\ref{ttnnmus}).
Blue filled circles are previous literature spectroscopic determinations, 
and the red filled triangle is a photometric measurement is the new
VLT/X-Shooter spectroscopic P-fix MCMC determination
(see Table~\ref{ttnnmus}).
The small panels show two MonteCarlo (MC) simulations of
10,000 realizations taking into account the uncertainties of each $T_n$ point: 
(i) with the observed data set, i.e. using $T_n$ values as a centre of the 
MC distributions (left small panel), and (ii) using the points on the 
parabolic fit (right small panel).
{\it Bottom panel}: residuals of the fit of the $T_n$ values versus 
the cycle number $n$.}
}
\label{fphnmus}   
\end{figure}

In Fig.~\ref{fphnmus} we have depicted the orbital phase
shift, defined as $\phi_n=\frac{T_n-T_0}{P_0}-n$, 
of each of the $T_n$ values as a function of the orbital cycle 
number $n$, together with the best-fitting second-order solution.     
This figure shows a clear deviation from the null variation and 
that $\dot P$ is negative. 
 
The result, $\dot P = -(6.6 \pm4.0) \times 10^{-10}$~s/s, 
which can also be expressed as $\dot P=-20.7\pm12.7$~ms~yr$^{-1}$, 
provides a determination at the $\sim1.6\sigma$ level 
(i.e. at $\sim 88$\% confidence).

Following \citet{gon14}, we perform two MonteCarlo (MC) simulations: 
(i) with the observed $T_n$ points by randomly varying their values 
with the uncertainties $\delta T_n$ in a normal distribution; 
(ii) with the fitted $T_n$ points on the parabolic fit.
We fit 10,000 realizations of the simulated $T_n$ points: for case (i) 
with weights in the fit given as $1/(\delta T_n)^2$; 
and for case (ii) without weights.
The resulting histograms of $\dot P$ are shown in two small panels 
within Fig.~\ref{fphnmus}.
 
These MC simulations give
$\dot P_{\rm MC} \sim -21$~ms~yr$^{-1}$ (with an uncertainty 
of 12.7 -- 14.2~ms~yr$^{-1}$) and confirms the extremely fast orbital 
shrinkage in \mbox{Nova Muscae 1991} (see Table~\ref{ttpar}).

\section{Discussion and conclusions}

Short period BHXBs exhibit negative orbital period derivative but at 
different rate, with \mbox{Nova Muscae 1991}  ($P_{\rm orb}\sim10.4$~hr) 
showing the fastest orbital decay.
Conventional models of AMLs due to GR, MB and ML are far from
being able to reproduce this behaviour. 
\citet{gon14} suggested an evolutionary sequence in which the orbital 
decays observed in SP-BHXBs would be faster as the 
companion star approaches the black hole. However, the case of 
\mbox{Nova Muscae 1991} rules out this hypothesis.
 
\citet{gon12a,gon14} estimated the orbital period 
derivative, $\dot P_{\rm MB,ML}$, when considering only AMLs due to 
MB and ML. They demonstrated 
that $\dot P_{\rm MB,ML}$ cannot account for the fast orbital decay observed 
in \mbox{XTE J1118+480}. Standard prescriptions of MB, 
assuming $\gamma =2.5$~\citep{ver93} and adopting arbitrarily 
$\beta=-\dot M_{\rm BH}/\dot M_2=0.5$ \citep{pod02a}, and the specific 
angular momentum, $j_w=1$, carried away by the mass lost from the system, 
provide values of $\dot P_{\rm MB,ML} \sim -0.028$~ms~yr$^{-1}$ for 
\mbox{Nova Muscae 1991}.
In the extreme and unrealistic case, $\gamma =0$, i.e the strongest 
possible MB effect, and $\beta=0$, i.e all the mass transferred by the 
secondary star is lost from the system, we only get 
$\dot P_{\rm MB,ML} \sim -0.11$~ms~yr$^{-1}$, which is 300 times 
smaller than the observed orbital period decay.
At orbital periods shorter than 3~hr, standard theory predicts that 
GR begins to dominate, but its contribution is totally 
negligible for \mbox{Nova Muscae 1991}, 
$\dot P_{\rm GR} \sim -0.011$~ms~yr$^{-1}$.

\citet{gon12a,gon14} suggested that extremely high magnetic fields at the
surface of the secondary star, in the range $B_S \sim 0.4$-$30$~kG, 
could help to explain the fast orbital decay in \mbox{XTE J1118+480} 
and \mbox{A0620-00}. The mass transfer rates~\citep{kin96b} of these 
two systems and \mbox{Nova Muscae 1991} are estimated to be 
$\dot M_2\sim$~0.10, 0.46, 1.67~$\times 10^{-9}$~\Msun yr$^{-1}$, respectively. 
Assuming conservative mass transfer (i.e. no mass lost from the system, 
$\dot M_{\rm BH} =  -\dot M_{\rm 2}$) and neglecting AML due to 
GR, we can make a rough estimate of the magnetic 
field, $B_S$, at the surface of the secondary required to accommodate the
period decays observed in these three BHXBs~\citep[see][for further details]{jus06}. 
We find $B_S\sim$ [0.7, 2.3, 7.4], [16, 50, 160], 
[12, 38, 120] kG, respectively, for three values of the wind-driving energy
efficiency factor, $f_\epsilon=$ [$10^{-1}$, $10^{-2}$, $10^{-3}$]~\citep{tav93}. 
The extremely high $B_S$ values estimated at low $f_\epsilon\sim10^{-3}$,
could be compensated with higher mass transfer rates. 
However, for $f_\epsilon=0.1$, the magnitudes of the surface magnetic fields
are still consistent with those in peculiar Ap stars~\citep{jus06}. 
As discussed in \citet{gon14} the secondary might have been able to retain 
the high magnetic field during the binary evolution.

Although speculative, these high magnetic fields might be connected 
with the (rotation induced) chromospheric activity on the companion star, 
as proposed for \mbox{A0620--00} \citep{gon10} to explain the observed 
H$\alpha$ emission feature of the secondary star.
H$\alpha$ emission from the companion has been also detected in 
\mbox{XTE J1118+480}~\citep{zur16} and \mbox{Nova Muscae 1991}
\citep{cas97} and (Gonz\'alez Hern\'andez et al. in preparation).

An alternative scenario is that we may be measuring orbital period 
modulations similar to what has been observed before in other type of 
binaries such as Algol objects, e.g. V471 Tau~\citep{ski88}, or cataclysmic 
variables~\citep[CVs][]{pri75,war88}. 
In these systems, orbital period modulations of the order of $\Delta P/P\sim10^{-5}-10^{-6}$ 
with different signs have been observed. Theoretical models invoking 
a variable gravitational quadrupole moment, due to internal deformations, 
produced by magnetic activity in the outer convection zone have been 
suggested as the mechanism responsible for those 
period changes~\citep{app87,app92}.
This mechanism would also imply extremely high magnetic fields~\citep{gon14}, 
but it appears quite unlikely that the large orbital period decays seen in 
these BHXBs (all with negative sign) are the result of orbital period modulations. 

On the other hand, \citet{sch16} have recently speculated that nova eruptions 
may produce frictional AML in CVs, as a possible solution of the white-dwarf 
(WD) mass problem, the orbital period distribution and the space density of CVs. 
This consequential AML could also occur in BHXBs during outburst with significant 
mass ejections, as has been very recently seen in V404 Cygni~\citep{mun16}.

Recently, \citet{che15} have proposed AML models including circumbinary discs
as a possible way to explain the fast orbital period decays in these BHXBs. This work
shows the evolution of orbital period and period derivative for \mbox{XTE J1118+480}
during $\sim0.8-1.2$~Gyr (for initial $M_2\sim 3-1.5$~\Msuno) of its binary evolution. 
The model requires a current mass transfer rate of about 
$10^{-8}$~\Msun yr$^{-1}$, which is significantly larger than the current 
estimated mass transfer rate of $10^{-10}$~\Msun yr$^{-1}$.
It also requires a circumbinary disc with a mass 
($M_{\rm CD}\sim2-3\times10^{-4}$~\Msuno) which is significantly larger than the 
inferred from mid-IR observations, i.e. $\sim10^{-9}$~\Msun~\citep{mun06}.
In addition, \citet{gal07} have suggested that non-thermal synchrotron 
emission from a jet could account for a significant fraction (or even all) of 
the measured excess mid-IR emission.

The orbital decays observed in these BHXBs, if kept constant, predict that the 
secondary star will reach the event horizon as given by the Schwarzschild radius 
of $\sim$22, 19 and 32 km in about 12~Myr for \mbox{XTE J1118+480}, 
70~Myr for \mbox{A0620--00}, and only 2.7~Myr for \mbox{Nova Muscae 1991}. 
This is an extremely short time-scale, although the companion's fate is probably 
not realistic. In a standard evolutionary scenario with AMLs driven by MB and 
GR, a minimum orbital period is expected~\citep[and observed in CVs, see][]{gan09}. 
The minimum period is caused by the response of the companion star 
to ML when it reaches the substellar limit~\citep{pac81}. 
It is unclear whether this scenario would be valid here but, in any case, the 
large orbital period decays measured clearly have very important implications 
on the evolution and lifetime of SP-BHXBs, which, according to the 
standard models, is typically $\sim5\times10^9$~yr. 
In addition, the companion stars with spectral types earlier than K5V show high 
Li abundances~\citep{mar96,gon04a,gon06}, similar to those of low mass stars
of the young (120~Myr) Pleiades cluster~\citep{gar94}, which are not easily explained 
unless a Li preservation mechanism is invoked, on the basis of high rotation 
velocities of the companion stars kept during the evolution of tidally 
locked BHXBs~\citep{mac05,cas07b}. 
A shorter lifetime, of $\sim10^7-10^8$~yr, would alleviate this high-Li problem
in BHXBs as well as the long-standing birthrate problem of millisecond pulsars 
and low-mass BHXBs~\citep{nay93,pod02a}.
Future observations of these BHXBs, in particular, of \mbox{Nova Muscae 1991}
will probably help to understand these extremely fast orbital period decays.

\section*{Acknowledgments}
JIGH acknowledges financial support from the Spanish Ministry of Economy 
and Competitiveness (MINECO) under the 2013 Ram\'on y Cajal program 
MINECO RYC-2013-14875, and JIGH and RRL also acknowledge financial 
support from the Spanish ministry project MINECO AYA2014-56359-P.
JC also acknowledges support from MINECO through the programme 
AYA2013-42627 and from the Leverhulme Trust through the Visiting 
Professorship Grant VP2-2015-046.

We are grateful to J. A. Rubi\~no Mart{\'\i}n for providing us with several
routines that allow us to implement the MCMC Bayesian analysis.

We are grateful to T. Marsh for the use of the MOLLY analysis package.
This work has made use of the IRAF facilities.




\bibliographystyle{mnras}
\bibliography{lmxbs}

\begin{thebibliography}{}
\makeatletter
\relax
\def\mn@urlcharsother{\let\do\@makeother \do\$\do\&\do\#\do\^\do\_\do\%\do\~}
\def\mn@doi{\begingroup\mn@urlcharsother \@ifnextchar [ {\mn@doi@}
  {\mn@doi@[]}}
\def\mn@doi@[#1]#2{\def\@tempa{#1}\ifx\@tempa\@empty \href
  {http://dx.doi.org/#2} {doi:#2}\else \href {http://dx.doi.org/#2} {#1}\fi
  \endgroup}
\def\mn@eprint#1#2{\mn@eprint@#1:#2::\@nil}
\def\mn@eprint@arXiv#1{\href {http://arxiv.org/abs/#1} {{\tt arXiv:#1}}}
\def\mn@eprint@dblp#1{\href {http://dblp.uni-trier.de/rec/bibtex/#1.xml}
  {dblp:#1}}
\def\mn@eprint@#1:#2:#3:#4\@nil{\def\@tempa {#1}\def\@tempb {#2}\def\@tempc
  {#3}\ifx \@tempc \@empty \let \@tempc \@tempb \let \@tempb \@tempa \fi \ifx
  \@tempb \@empty \def\@tempb {arXiv}\fi \@ifundefined
  {mn@eprint@\@tempb}{\@tempb:\@tempc}{\expandafter \expandafter \csname
  mn@eprint@\@tempb\endcsname \expandafter{\@tempc}}}

\bibitem[\protect\citeauthoryear{{Applegate}}{{Applegate}}{1992}]{app92}
{Applegate} J.~H.,  1992, \mn@doi [\apj] {10.1086/170967}, \href
  {http://adsabs.harvard.edu/abs/1992ApJ...385..621A} {385, 621}

\bibitem[\protect\citeauthoryear{{Applegate} \& {Patterson}}{{Applegate} \&
  {Patterson}}{1987}]{app87}
{Applegate} J.~H.,  {Patterson} J.,  1987, \mn@doi [\apjl] {10.1086/185044},
  \href {http://adsabs.harvard.edu/abs/1987ApJ...322L..99A} {322, L99}

\bibitem[\protect\citeauthoryear{{Calvelo}, {Vrtilek}, {Steeghs}, {Torres},
  {Neilsen}, {Filippenko}  \& {Gonz{\'a}lez Hern{\'a}ndez}}{{Calvelo}
  et~al.}{2009}]{cal09}
{Calvelo} D.~E.,  {Vrtilek} S.~D.,  {Steeghs} D.,  {Torres} M.~A.~P.,
  {Neilsen} J.,  {Filippenko} A.~V.,   {Gonz{\'a}lez Hern{\'a}ndez} J.~I.,
  2009, \mn@doi [\mnras] {10.1111/j.1365-2966.2009.15304.x}, \href
  {http://adsabs.harvard.edu/abs/2009MNRAS.399..539C} {399, 539}

\bibitem[\protect\citeauthoryear{{Cantrell} et~al.,}{{Cantrell}
  et~al.}{2010}]{can10}
{Cantrell} A.~G.,  et~al., 2010, \mn@doi [\apj] {10.1088/0004-637X/710/2/1127},
  \href {http://adsabs.harvard.edu/abs/2010ApJ...710.1127C} {710, 1127}

\bibitem[\protect\citeauthoryear{{Casares}, {Mart{\'{\i}}n}, {Charles},
  {Molaro}  \& {Rebolo}}{{Casares} et~al.}{1997}]{cas97}
{Casares} J.,  {Mart{\'{\i}}n} E.~L.,  {Charles} P.~A.,  {Molaro} P.,
  {Rebolo} R.,  1997, \mn@doi [\na] {10.1016/S1384-1076(96)00022-X}, \href
  {http://adsabs.harvard.edu/abs/1997NewA....1..299C} {1, 299}

\bibitem[\protect\citeauthoryear{{Casares}, {Bonifacio}, {Gonz{\'a}lez
  Hern{\'a}ndez}, {Molaro}  \& {Zoccali}}{{Casares} et~al.}{2007}]{cas07b}
{Casares} J.,  {Bonifacio} P.,  {Gonz{\'a}lez Hern{\'a}ndez} J.~I.,  {Molaro}
  P.,   {Zoccali} M.,  2007, \mn@doi [\aap] {10.1051/0004-6361:20066875}, \href
  {http://adsabs.harvard.edu/abs/2007A%26A...470.1033C} {470, 1033}

\bibitem[\protect\citeauthoryear{{Chen} \& {Li}}{{Chen} \& {Li}}{2015}]{che15}
{Chen} W.-C.,  {Li} X.-D.,  2015, \mn@doi [\aap] {10.1051/0004-6361/201526524},
  \href {http://adsabs.harvard.edu/abs/2015A%26A...583A.108C} {583, A108}

\bibitem[\protect\citeauthoryear{{Gallo}, {Migliari}, {Markoff}, {Tomsick},
  {Bailyn}, {Berta}, {Fender}  \& {Miller-Jones}}{{Gallo} et~al.}{2007}]{gal07}
{Gallo} E.,  {Migliari} S.,  {Markoff} S.,  {Tomsick} J.~A.,  {Bailyn} C.~D.,
  {Berta} S.,  {Fender} R.,   {Miller-Jones} J.~C.~A.,  2007, \mn@doi [\apj]
  {10.1086/521524}, \href {http://adsabs.harvard.edu/abs/2007ApJ...670..600G}
  {670, 600}

\bibitem[\protect\citeauthoryear{{G{\"a}nsicke} et~al.,}{{G{\"a}nsicke}
  et~al.}{2009}]{gan09}
{G{\"a}nsicke} B.~T.,  et~al., 2009, \mn@doi [\mnras]
  {10.1111/j.1365-2966.2009.15126.x}, \href
  {http://adsabs.harvard.edu/abs/2009MNRAS.397.2170G} {397, 2170}

\bibitem[\protect\citeauthoryear{{Garcia Lopez}, {Rebolo}  \& {Martin}}{{Garcia
  Lopez} et~al.}{1994}]{gar94}
{Garcia Lopez} R.~J.,  {Rebolo} R.,   {Martin} E.~L.,  1994, \aap, \href
  {http://adsabs.harvard.edu/abs/1994A%26A...282..518G} {282, 518}

\bibitem[\protect\citeauthoryear{{Gelino}}{{Gelino}}{2004}]{gel04}
{Gelino} D.~M.,  2004, in {Tovmassian} G.,  {Sion} E.,  eds,  Vol. 20, Revista
  Mexicana de Astronomia y Astrofisica Conference Series. pp 214--214

\bibitem[\protect\citeauthoryear{{Gonz{\'a}lez Hern{\'a}ndez} \&
  {Casares}}{{Gonz{\'a}lez Hern{\'a}ndez} \& {Casares}}{2010}]{gon10}
{Gonz{\'a}lez Hern{\'a}ndez} J.~I.,  {Casares} J.,  2010, \mn@doi [\aap]
  {10.1051/0004-6361/201014088}, \href
  {http://adsabs.harvard.edu/abs/2010A%26A...516A..58G} {516, A58}

\bibitem[\protect\citeauthoryear{{Gonz{\'a}lez Hern{\'a}ndez}, {Rebolo},
  {Israelian}, {Casares}, {Maeder}  \& {Meynet}}{{Gonz{\'a}lez Hern{\'a}ndez}
  et~al.}{2004}]{gon04a}
{Gonz{\'a}lez Hern{\'a}ndez} J.~I.,  {Rebolo} R.,  {Israelian} G.,  {Casares}
  J.,  {Maeder} A.,   {Meynet} G.,  2004, \mn@doi [\apj] {10.1086/421102},
  \href {http://adsabs.harvard.edu/abs/2004ApJ...609..988G} {609, 988}

\bibitem[\protect\citeauthoryear{{Gonz{\'a}lez Hern{\'a}ndez}, {Rebolo},
  {Israelian}, {Harlaftis}, {Filippenko}  \& {Chornock}}{{Gonz{\'a}lez
  Hern{\'a}ndez} et~al.}{2006}]{gon06}
{Gonz{\'a}lez Hern{\'a}ndez} J.~I.,  {Rebolo} R.,  {Israelian} G.,  {Harlaftis}
  E.~T.,  {Filippenko} A.~V.,   {Chornock} R.,  2006, \mn@doi [\apjl]
  {10.1086/505391}, \href {http://adsabs.harvard.edu/abs/2006ApJ...644L..49G}
  {644, L49}

\bibitem[\protect\citeauthoryear{{Gonz{\'a}lez Hern{\'a}ndez}, {Rebolo},
  {Israelian}, {Filippenko}, {Chornock}, {Tominaga}, {Umeda}  \&
  {Nomoto}}{{Gonz{\'a}lez Hern{\'a}ndez} et~al.}{2008}]{gon08b}
{Gonz{\'a}lez Hern{\'a}ndez} J.~I.,  {Rebolo} R.,  {Israelian} G.,
  {Filippenko} A.~V.,  {Chornock} R.,  {Tominaga} N.,  {Umeda} H.,   {Nomoto}
  K.,  2008, \mn@doi [\apj] {10.1086/586888}, \href
  {http://adsabs.harvard.edu/abs/2008ApJ...679..732G} {679, 732}

\bibitem[\protect\citeauthoryear{{Gonz{\'a}lez Hern{\'a}ndez}, {Casares},
  {Rebolo}, {Israelian}, {Filippenko}  \& {Chornock}}{{Gonz{\'a}lez
  Hern{\'a}ndez} et~al.}{2011}]{gon11}
{Gonz{\'a}lez Hern{\'a}ndez} J.~I.,  {Casares} J.,  {Rebolo} R.,  {Israelian}
  G.,  {Filippenko} A.~V.,   {Chornock} R.,  2011, \mn@doi [\apj]
  {10.1088/0004-637X/738/1/95}, \href
  {http://adsabs.harvard.edu/abs/2011ApJ...738...95G} {738, 95}

\bibitem[\protect\citeauthoryear{{Gonz{\'a}lez Hern{\'a}ndez}, {Rebolo}  \&
  {Casares}}{{Gonz{\'a}lez Hern{\'a}ndez} et~al.}{2012}]{gon12a}
{Gonz{\'a}lez Hern{\'a}ndez} J.~I.,  {Rebolo} R.,   {Casares} J.,  2012,
  \mn@doi [\apjl] {10.1088/2041-8205/744/2/L25}, \href
  {http://adsabs.harvard.edu/abs/2012ApJ...744L..25G} {744, L25}

\bibitem[\protect\citeauthoryear{{Gonz{\'a}lez Hern{\'a}ndez}, {Rebolo}  \&
  {Casares}}{{Gonz{\'a}lez Hern{\'a}ndez} et~al.}{2014}]{gon14}
{Gonz{\'a}lez Hern{\'a}ndez} J.~I.,  {Rebolo} R.,   {Casares} J.,  2014,
  \mn@doi [\mnras] {10.1093/mnrasl/slt150}, \href
  {http://adsabs.harvard.edu/abs/2014MNRAS.438L..21G} {438, L21}

\bibitem[\protect\citeauthoryear{{Ivanova}}{{Ivanova}}{2006}]{iva06b}
{Ivanova} N.,  2006, \mn@doi [\apjl] {10.1086/510672}, \href
  {http://adsabs.harvard.edu/abs/2006ApJ...653L.137I} {653, L137}

\bibitem[\protect\citeauthoryear{{Justham}, {Rappaport}  \&
  {Podsiadlowski}}{{Justham} et~al.}{2006}]{jus06}
{Justham} S.,  {Rappaport} S.,   {Podsiadlowski} P.,  2006, \mn@doi [\mnras]
  {10.1111/j.1365-2966.2005.09907.x}, \href
  {http://adsabs.harvard.edu/abs/2006MNRAS.366.1415J} {366, 1415}

\bibitem[\protect\citeauthoryear{{Khargharia}, {Froning}, {Robinson}  \&
  {Gelino}}{{Khargharia} et~al.}{2013}]{kha13}
{Khargharia} J.,  {Froning} C.~S.,  {Robinson} E.~L.,   {Gelino} D.~M.,  2013,
  \mn@doi [\aj] {10.1088/0004-6256/145/1/21}, \href
  {http://adsabs.harvard.edu/abs/2013AJ....145...21K} {145, 21}

\bibitem[\protect\citeauthoryear{{King}, {Kolb}  \& {Burderi}}{{King}
  et~al.}{1996}]{kin96b}
{King} A.~R.,  {Kolb} U.,   {Burderi} L.,  1996, \mn@doi [\apjl]
  {10.1086/310105}, \href {http://adsabs.harvard.edu/abs/1996ApJ...464L.127K}
  {464, L127}

\bibitem[\protect\citeauthoryear{{Landau} \& {Lifshitz}}{{Landau} \&
  {Lifshitz}}{1962}]{lan62}
{Landau} L.~D.,  {Lifshitz} E.~M.,  1962, {The Classical Theory of Fields.
  Pergamon, Oxford}

\bibitem[\protect\citeauthoryear{{Maccarone}, {Jonker}  \& {Sills}}{{Maccarone}
  et~al.}{2005}]{mac05}
{Maccarone} T.~J.,  {Jonker} P.~G.,   {Sills} A.~I.,  2005, \mn@doi [\aap]
  {10.1051/0004-6361:20052791}, \href
  {http://adsabs.harvard.edu/abs/2005A%26A...436..671M} {436, 671}

\bibitem[\protect\citeauthoryear{{Mart{\'{\i}}n}, {Casares}, {Molaro}, {Rebolo}
   \& {Charles}}{{Mart{\'{\i}}n} et~al.}{1996}]{mar96}
{Mart{\'{\i}}n} E.~L.,  {Casares} J.,  {Molaro} P.,  {Rebolo} R.,   {Charles}
  P.,  1996, \mn@doi [\na] {10.1016/S1384-1076(96)00014-0}, \href
  {http://adsabs.harvard.edu/abs/1996NewA....1..197M} {1, 197}

\bibitem[\protect\citeauthoryear{{McClintock} \& {Remillard}}{{McClintock} \&
  {Remillard}}{1986}]{mcc86}
{McClintock} J.~E.,  {Remillard} R.~A.,  1986, \mn@doi [\apj] {10.1086/164482},
  \href {http://adsabs.harvard.edu/abs/1986ApJ...308..110M} {308, 110}

\bibitem[\protect\citeauthoryear{{Mu{\~n}oz-Darias} et~al.,}{{Mu{\~n}oz-Darias}
  et~al.}{2016}]{mun16}
{Mu{\~n}oz-Darias} T.,  et~al., 2016, \mn@doi [\nat] {10.1038/nature17446},
  \href {http://adsabs.harvard.edu/abs/2016Natur.534...75M} {534, 75}

\bibitem[\protect\citeauthoryear{{Muno} \& {Mauerhan}}{{Muno} \&
  {Mauerhan}}{2006}]{mun06}
{Muno} M.~P.,  {Mauerhan} J.,  2006, \mn@doi [\apjl] {10.1086/507990}, \href
  {http://adsabs.harvard.edu/abs/2006ApJ...648L.135M} {648, L135}

\bibitem[\protect\citeauthoryear{{Naylor} \& {Podsiadlowski}}{{Naylor} \&
  {Podsiadlowski}}{1993}]{nay93}
{Naylor} T.,  {Podsiadlowski} P.,  1993, \mn@doi [\mnras]
  {10.1093/mnras/262.4.929}, \href
  {http://adsabs.harvard.edu/abs/1993MNRAS.262..929N} {262, 929}

\bibitem[\protect\citeauthoryear{{Neilsen}, {Steeghs}  \& {Vrtilek}}{{Neilsen}
  et~al.}{2008}]{nei08}
{Neilsen} J.,  {Steeghs} D.,   {Vrtilek} S.~D.,  2008, \mn@doi [\mnras]
  {10.1111/j.1365-2966.2007.12599.x}, \href
  {http://adsabs.harvard.edu/abs/2008MNRAS.384..849N} {384, 849}

\bibitem[\protect\citeauthoryear{{Orosz}, {Bailyn}, {McClintock}  \&
  {Remillard}}{{Orosz} et~al.}{1996}]{oro96}
{Orosz} J.~A.,  {Bailyn} C.~D.,  {McClintock} J.~E.,   {Remillard} R.~A.,
  1996, \mn@doi [\apj] {10.1086/177698}, \href
  {http://adsabs.harvard.edu/abs/1996ApJ...468..380O} {468, 380}

\bibitem[\protect\citeauthoryear{{Paczynski}}{{Paczynski}}{1981}]{pac81}
{Paczynski} B.,  1981, \actaa, \href
  {http://adsabs.harvard.edu/abs/1981AcA....31....1P} {31, 1}

\bibitem[\protect\citeauthoryear{{Podsiadlowski}, {Rappaport}  \&
  {Pfahl}}{{Podsiadlowski} et~al.}{2002}]{pod02a}
{Podsiadlowski} P.,  {Rappaport} S.,   {Pfahl} E.~D.,  2002, \mn@doi [\apj]
  {10.1086/324686}, \href {http://adsabs.harvard.edu/abs/2002ApJ...565.1107P}
  {565, 1107}

\bibitem[\protect\citeauthoryear{{Pringle}}{{Pringle}}{1975}]{pri75}
{Pringle} J.~E.,  1975, \mnras, \href
  {http://adsabs.harvard.edu/abs/1975MNRAS.170..633P} {170, 633}

\bibitem[\protect\citeauthoryear{{Rappaport}, {Joss}  \& {Webbink}}{{Rappaport}
  et~al.}{1982}]{rap82}
{Rappaport} S.,  {Joss} P.~C.,   {Webbink} R.~F.,  1982, \mn@doi [\apj]
  {10.1086/159772}, \href {http://adsabs.harvard.edu/abs/1982ApJ...254..616R}
  {254, 616}

\bibitem[\protect\citeauthoryear{{Remillard}, {McClintock}  \&
  {Bailyn}}{{Remillard} et~al.}{1992}]{rem92}
{Remillard} R.~A.,  {McClintock} J.~E.,   {Bailyn} C.~D.,  1992, \mn@doi
  [\apjl] {10.1086/186628}, \href
  {http://adsabs.harvard.edu/abs/1992ApJ...399L.145R} {399, L145}

\bibitem[\protect\citeauthoryear{{Rubi{\~n}o-Martin}
  et~al.,}{{Rubi{\~n}o-Martin} et~al.}{2003}]{rub03}
{Rubi{\~n}o-Martin} J.~A.,  et~al., 2003, \mn@doi [\mnras]
  {10.1046/j.1365-8711.2003.06494.x}, \href
  {http://adsabs.harvard.edu/abs/2003MNRAS.341.1084R} {341, 1084}

\bibitem[\protect\citeauthoryear{{Schreiber}, {Zorotovic}  \&
  {Wijnen}}{{Schreiber} et~al.}{2016}]{sch16}
{Schreiber} M.~R.,  {Zorotovic} M.,   {Wijnen} T.~P.~G.,  2016, \mn@doi
  [\mnras] {10.1093/mnrasl/slv144}, \href
  {http://adsabs.harvard.edu/abs/2016MNRAS.455L..16S} {455, L16}

\bibitem[\protect\citeauthoryear{{Shahbaz}, {Naylor}  \& {Charles}}{{Shahbaz}
  et~al.}{1997}]{sha97}
{Shahbaz} T.,  {Naylor} T.,   {Charles} P.~A.,  1997, \mn@doi [\mnras]
  {10.1093/mnras/285.3.607}, \href
  {http://adsabs.harvard.edu/abs/1997MNRAS.285..607S} {285, 607}

\bibitem[\protect\citeauthoryear{{Skillman} \& {Patterson}}{{Skillman} \&
  {Patterson}}{1988}]{ski88}
{Skillman} D.~R.,  {Patterson} J.,  1988, \mn@doi [\aj] {10.1086/114857}, \href
  {http://adsabs.harvard.edu/abs/1988AJ.....96..976S} {96, 976}

\bibitem[\protect\citeauthoryear{{Tavani} \& {London}}{{Tavani} \&
  {London}}{1993}]{tav93}
{Tavani} M.,  {London} R.,  1993, \mn@doi [\apj] {10.1086/172744}, \href
  {http://adsabs.harvard.edu/abs/1993ApJ...410..281T} {410, 281}

\bibitem[\protect\citeauthoryear{{Taylor} \& {Weisberg}}{{Taylor} \&
  {Weisberg}}{1982}]{tay82}
{Taylor} J.~H.,  {Weisberg} J.~M.,  1982, \mn@doi [\apj] {10.1086/159690},
  \href {http://adsabs.harvard.edu/abs/1982ApJ...253..908T} {253, 908}

\bibitem[\protect\citeauthoryear{{Verbunt}}{{Verbunt}}{1993}]{ver93}
{Verbunt} F.,  1993, \mn@doi [\araa] {10.1146/annurev.aa.31.090193.000521},
  \href {http://adsabs.harvard.edu/abs/1993ARA%26A..31...93V} {31, 93}

\bibitem[\protect\citeauthoryear{{Verbunt} \& {Zwaan}}{{Verbunt} \&
  {Zwaan}}{1981}]{ver81}
{Verbunt} F.,  {Zwaan} C.,  1981, \aap, \href
  {http://adsabs.harvard.edu/abs/1981A%26A...100L...7V} {100, L7}

\bibitem[\protect\citeauthoryear{{Vernet}, {Dekker}, {D'Odorico}, {Kaper},
  {Kjaergaard}, {Hammer}, {Randich}  \& {Zerbi}}{{Vernet} et~al.}{2011}]{ver11}
{Vernet} J.,  {Dekker} H.,  {D'Odorico} S.,  {Kaper} L.,  {Kjaergaard} P.,
  {Hammer} F.,  {Randich} S.,   {Zerbi} F.,  2011, \mn@doi [\aap]
  {10.1051/0004-6361/201117752}, \href
  {http://adsabs.harvard.edu/abs/2011A%26A...536A.105V} {536, A105}

\bibitem[\protect\citeauthoryear{{Wang} \& {Wang}}{{Wang} \&
  {Wang}}{2014}]{wan14}
{Wang} X.,  {Wang} Z.,  2014, \mn@doi [\apj] {10.1088/0004-637X/788/2/184},
  \href {http://adsabs.harvard.edu/abs/2014ApJ...788..184W} {788, 184}

\bibitem[\protect\citeauthoryear{{Warner}}{{Warner}}{1988}]{war88}
{Warner} B.,  1988, \mn@doi [\nat] {10.1038/336129a0}, \href
  {http://adsabs.harvard.edu/abs/1988Natur.336..129W} {336, 129}

\bibitem[\protect\citeauthoryear{{Wu} et~al.,}{{Wu} et~al.}{2015}]{wu15}
{Wu} J.,  et~al., 2015, \mn@doi [\apj] {10.1088/0004-637X/806/1/92}, \href
  {http://adsabs.harvard.edu/abs/2015ApJ...806...92W} {806, 92}

\bibitem[\protect\citeauthoryear{{Wu}, {Orosz}, {McClintock}, {Hasan},
  {Bailyn}, {Gou}  \& {Chen}}{{Wu} et~al.}{2016}]{wu16}
{Wu} J.,  {Orosz} J.~A.,  {McClintock} J.~E.,  {Hasan} I.,  {Bailyn} C.~D.,
  {Gou} L.,   {Chen} Z.,  2016, \mn@doi [\apj] {10.3847/0004-637X/825/1/46},
  \href {http://adsabs.harvard.edu/abs/2016ApJ...825...46W} {825, 46}

\bibitem[\protect\citeauthoryear{{Yagi}}{{Yagi}}{2012}]{yag12}
{Yagi} K.,  2012, \mn@doi [\prd] {10.1103/PhysRevD.86.081504}, \href
  {http://adsabs.harvard.edu/abs/2012PhRvD..86h1504Y} {86, 081504}

\bibitem[\protect\citeauthoryear{{Yungelson} \& {Lasota}}{{Yungelson} \&
  {Lasota}}{2008}]{yun08b}
{Yungelson} L.~R.,  {Lasota} J.-P.,  2008, \mn@doi [\aap]
  {10.1051/0004-6361:200809684}, \href
  {http://adsabs.harvard.edu/abs/2008A%26A...488..257Y} {488, 257}

\bibitem[\protect\citeauthoryear{{Zurita}, {Gonz{\'a}lez Hern{\'a}ndez},
  {Escorza}  \& {Casares}}{{Zurita} et~al.}{2016}]{zur16}
{Zurita} C.,  {Gonz{\'a}lez Hern{\'a}ndez} J.~I.,  {Escorza} A.,   {Casares}
  J.,  2016, \mn@doi [\mnras] {10.1093/mnras/stw1235}, \href
  {http://adsabs.harvard.edu/abs/2016MNRAS.460.4289Z} {460, 4289}

\makeatother
\end{thebibliography}






\bsp	
\label{lastpage}
\end{document}